\documentclass[prd,twocolumn,nofootinbib]{revtex4}
\usepackage{amsmath,amsfonts,amssymb}
\usepackage{hyperref,graphicx,dcolumn}

\newcommand{\be}{\begin{equation}}
\newcommand{\ee}{\end{equation}}
\newcommand{\bea}{\begin{eqnarray}}
\newcommand{\eea}{\end{eqnarray}}

\newcommand{\aap}{Astron. Astrophys.}
\newcommand{\aj}{Astron. J.}
\newcommand{\apjl}{Astrophys. J. Lett.}
\newcommand{\apjs}{Astrophys. J. Suppl. Ser.}
\newcommand{\araa}{Annu. Rev. Astron. Astrophys.}
\newcommand{\mnras}{Mon. Not. R. Astron. Soc.}
\newcommand{\physrep}{Phys. Rep.}
\begin{document}

\title{How Tomographic Cosmic Shear Maps Lead to Constraints on Dark Energy Properties}

\author{Hu Zhan}
\email{zhan@physics.ucdavis.edu}
\author{Lloyd Knox} 
\email{lknox@ucdavis.edu}
\affiliation{Department of Physics, University of California, Davis, CA 
             95616}

\begin{abstract}
  Using a number of numerical tests and analytic arguments we
  investigate {\em how} measurements of cosmic shear lead to
  constraints on dark energy.  We find that, in contrast to the case
  with galaxy number density correlation functions, standard rulers in
  the matter power spectrum play no significant role.  Sensitivity to
  distance ratios is provided by the ratios in the lensing kernel.  An
  absolute distance scale can only be established by breaking a
  potential degeneracy between growth and distance which can be done
  if the growth-redshift relation and distance-redshift relations are
  parameterized with sufficiently few parameters.  For the quality of
  dark energy determination, growth determination is primarily
  important because it improves the distance reconstructions.  The
  information about dark energy in the growth-redshift relation is
  always of secondary importance though the amount it contributes is
  highly dependent on what priors are taken in the cosmological
  parameter space.  We also explain the dependence of dark energy
  constraints from cosmic shear, relative distance measures
  (supernovae) and absolute distance measures (baryon acoustic
  oscillations) on assumptions about the mean curvature.
\end{abstract}


\maketitle

\section{Introduction}\label{sec:intr}

The manner by which galaxy number density two-point correlation
functions can be used to constrain dark energy is quite straightforward
and well understood.  In brief, the feature in the correlation function
that arises from acoustic oscillations in the pre-recombination baryon-
photon fluid is a standard ruler calibrated by cosmic microwave
background (CMB) observations \cite{peebles70, bond84, holtzman89}.  
Detection of the
angular extent of this feature (and/or redshift extent) at different
redshifts allows for determination of the angular-diameter distance
as a function of redshift, $D_A(z)$, (and/or expansion rate as a function
of redshift, $H(z)$).  The constraints on dark energy then follow
from the dependence of these quantities on dark energy parameters.
Similar statements could be made about the degree to which we
understand the use of standard candles such as supernovae or
standard populations such as clusters of galaxies --- even though
the latter is indeed a more complicated probe due to the dependence
of the population statistics on the rate of growth $G(z)$ of 
large-scale structure.

In contrast, the manner by which cosmic shear two-point correlation
functions constrain dark energy is not as straightforward nor is it as
well understood.  Others
\citep{abazajian03,simpson05,zhang05,knox06b} have already addressed
this subject.  Here we consider the question further.  Whereas in
\cite{song05} and \cite{knox06b}
it was shown that one can reconstruct the
distance-redshift and growth-redshift relations from cosmic shear
data, here we explain {\em how} these functions are reconstructed.
Whereas in \cite{zhang05} it was shown how information about relative
distances could be extracted from cosmic shear data in a manner
independent of assumptions about the matter power spectrum and its
growth over time, here we show how an {\em absolute} distance scale
can be reconstructed as well, and discuss the model-dependence of the
reconstruction.  As in \cite{abazajian03,simpson05,zhang05,knox06b}
we comment further on the different roles played by growth-redshift
and distance-redshift in constraining dark energy.

Baryon acoustic oscillations (BAOs)
\citep[e.g.,][]{eisenstein98, blake03,hu03b,linder03,seo03} and weak
lensing (WL)
\cite[e.g.,][]{hu99a,mellier99,bartelmann01,huterer02,refregier03,song04}
are emerging as two promising probes of dark
energy. Recent observations have detected the cosmic shear
\cite{bacon00,wittman00,vanwaerbeke01,jarvis03,hoekstra05} and baryon
wiggles \cite{cole05,eisenstein05,blake06b,huetsi06,padmanabhan06},
and they have been used to determine cosmological parameters such as
the matter density and normalization of the matter power spectrum.

We know that WL surveys can constrain dark energy through the dependence
of the distance-redshift relation and the growth-redshift relation
on dark energy.  Thus we begin by examining how these functions
change as cosmological parameters are varied.  We do so for both
relative distance measures ($D_A(z)H_0$) and absolute distance
measures ($D_A(z)$).  Comparison between these two leads to
an explanation of why Type Ia supernovae (SNe) are much more sensitive 
to assumptions about mean spatial curvature than is the case for BAOs.

In our effort to understand how WL constrains dark energy, we have
found it useful to compare WL with what happens in the galaxy
two-point correlation function case.  There are two important
differences between galaxy two-point correlation functions and shear
two-point correlation functions.  Shear two-point correlation
functions do not depend on any unknown bias factor, and do depend on
the ratio of distances in the lensing kernel. 
We find that both differences have a significant impact on 
dark energy constraints.

To assess the impact of the growth information
obtainable due to the lack of an unknown bias factor, the impact of
the distance dependence arising from the lensing kernel and various
features in the power spectrum, we carry out a number of tests built
upon a scale-free matter power spectrum and the normal cold dark
matter (CDM) power spectrum with or without the baryon wiggles. We
caution the reader that many systematic uncertainties are neglected in
these tests in order to isolate the effect of the subject under
investigation.  Because of this, the results of these tests should not
be taken as forecasts for future surveys.

One of the useful conclusions from our numerical tests is that power
spectrum features play no significant role in WL-derived dark energy
constraints, at least in the limit of infinite source density, i.e.,
no shot (shape) noise.  We can thus discuss the response of shear
power spectra to $D_A(z)$ and $G(z)$ using analytic expressions that
simplify greatly for the case of power-law power spectra.  In the
presence of noise, changes to the power spectrum shape due to
non-linear evolution boost the signal-to-noise ratio and thereby do
lead to tighter constraints.  For BAO, the nonlinear feature in the
galaxy/matter power spectra can be a potential standard ruler for
measuring the angular diameter distance. However, to take advantage of
this prominent but evolving feature, one must model the nonlinearity
and scale-dependent galaxy bias accurately.

We find that the sensitivity of WL to growth is mostly a nuisance.
That is, the information about dark energy that can be inferred
from constraints on the growth factor is subdominant to the dark
energy information that can be inferred from distance constraints.
The determination of the growth factor though, or more directly the
amplitude of the power spectrum as a function of time, is important
because it improves the distance determinations.  

Throughout this paper, we assume a low-density CDM universe with the 
following parameters: the dark energy equation-of-state parameters 
$w_0$ and $w_a$ as defined by $w(a) = w_0 + w_a(1-a)$, the matter 
density $\omega_{\rm m}$, the baryon density $\omega_{\rm b}$, the 
angular size of the sound horizon at the last scattering surface 
$\theta_{\rm s}$, the equivalent matter fraction of curvature 
$\Omega_{\rm K}$, the optical depth to scattering by electrons in the 
reionized inter-galactic medium, $\tau$, the primordial helium mass 
fraction $Y_{\rm p}$, the spectral index $n_{\rm s}$ and running 
$\alpha_{\rm s}$ of the primordial 
scalar perturbation power spectrum, and the normalization of the 
primordial curvature power spectrum $\Delta_R^2$ at 
$k = 0.05\,\mbox{Mpc}^{-1}$. The fiducial values are taken from 
the 3-year {\it WMAP} data \cite{spergel06}: 
($w_0$, $w_a$, $\omega_{\rm m}$, $\omega_{\rm b}$, $\theta_{\rm s}$,
$\Omega_{\rm K}$, $\tau$, $Y_{\rm p}$, $n_{\rm s}$, $\alpha_{\rm s}$,
$\Delta_R^2$) = 
(-1, 0, 0.127, 0.0223, 0.596$^\circ$, 0, 0.09, 0.24, 0.951, 0,
$2.0 \times 10^{-9}$). The reduced Hubble constant $h = 0.73$ and 
the present equivalent matter fraction of dark energy 
$\Omega_{\rm X} = 0.76$ are implicit in our parametrization. 

The rest of the paper is organized as follows. \S \ref{sec:dg} 
examines the utility and reconstruction of distance and growth for
studying dark energy. In \S \ref{sec:test}, we 
compare cosmological constraints from BAO and WL obtained with 
various features in the matter power spectrum. We discuss the 
results and conclude in \S \ref{sec:con}.

\begin{figure*}
\centering
\includegraphics[height=2.8in]{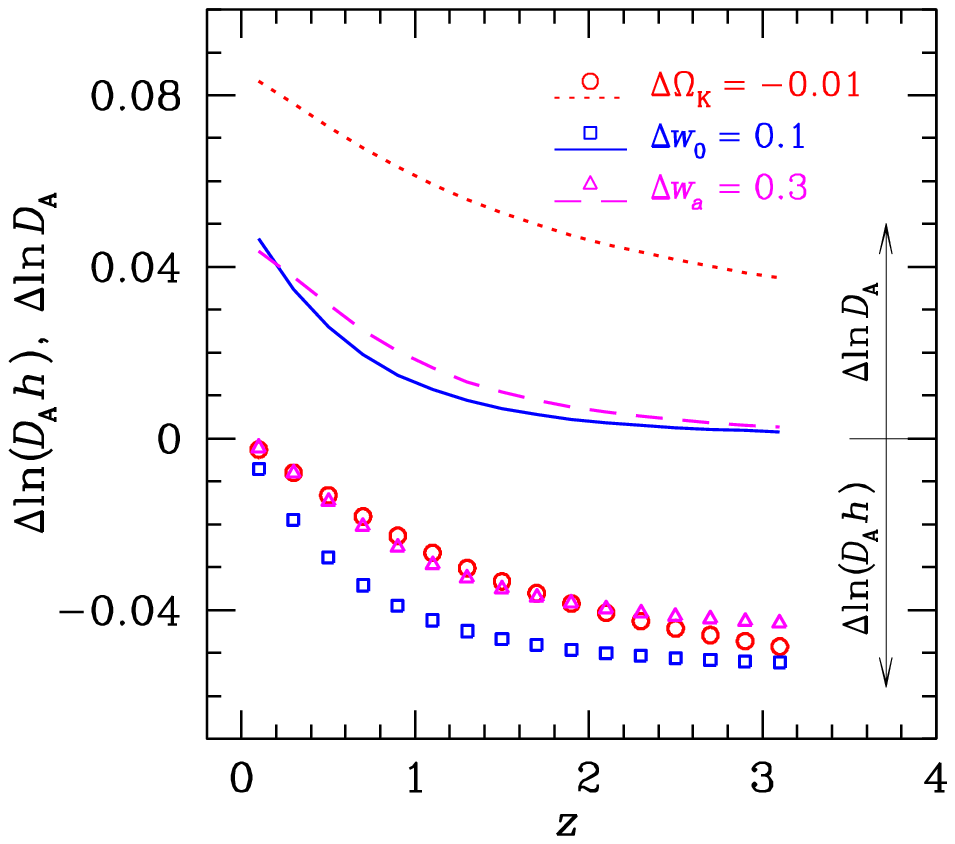} \hspace{0.2in}
\includegraphics[height=2.8in]{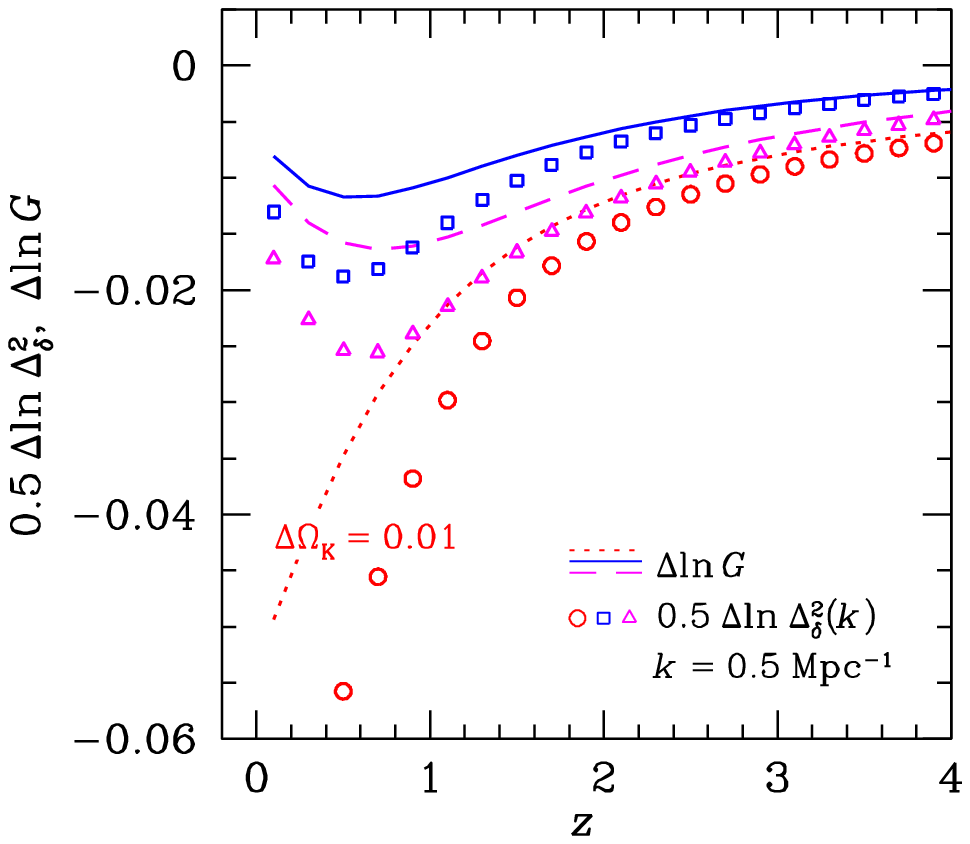}
\caption[f1]{ {\it Left panel}:
Fractional changes of the absolute comoving angular diameter distance 
$D_{\rm A}$ (lines) and the relative distance $D_{\rm A} h$ (symbols)
with respect to 
a small deviation of the dark energy EOS parameters $w_0$ and $w_a$ 
and the mean curvature parameter $\Omega_{\rm K}$ from their fiducial 
values. The fractional changes are calculated with all other 
parameters fixed. Fractional changes of the implicit Hubble constant 
equal $\Delta \ln D_{\rm A}$ at $z \sim 0$
 or $\Delta \ln (D_{\rm A}h)$ at recombination.
{\it Right panel}: 
Same as the left panel but for the linear growth function $G$
(lines) and nonlinear growth at $k = 0.5\,\mbox{Mpc}^{-1}$ (symbols).
The linear growth function follows the convention that 
$G(a) = a$ in an Einstein-de Sitter universe but $G(a=1)\ne 1$ in 
general. The nonlinear matter power 
spectrum is obtained by applying the \citet{peacock96} correction to
the no-wiggle CDM power spectrum \cite{eisenstein99a}.
\label{fig:dddg}}
\end{figure*}

\section{Probing Dark Energy with Distance and Growth} 
\label{sec:dg}

The relative strength of dark energy constraints from distance and 
growth is determined by two factors: the sensitivity to dark energy
parameters and the precision with which they can be reconstructed. 
Since distance is more sensitive to the parameters 
\cite{zhan06d} and since growth is determined several 
times worse 
than distance \cite{knox06b}, the WL constraints on dark energy 
from the reconstructed distances are considerably stronger than 
those from the linear growth function \cite{abazajian03,knox06b}. 
We note though that this conclusion is not reached consistently.
Both \cite{simpson05} and \cite{zhang05} find the dark energy
constraints from growth to be comparable to those
from distance, as we will discuss below.  

Distance and growth measurements in WL are entangled.  
For example, the precision on distance
depends on how well one can measure the growth function. 
Hence, even if the growth function does not place very powerful 
constraints directly, being able to measure it well can
still be crucial for probing dark energy.

\subsection{Sensitivity}

We show in Fig.~\ref{fig:dddg} fractional changes of distance (left
panel) and growth (right panel) with respect to a small deviation of 
the dark energy EOS parameters $w_0$ and $w_a$ and the mean curvature 
parameter $\Omega_{\rm K}$ from their fiducial values 
\cite[see also][]{hu05, zhan06d}. The fractional changes are 
calculated with all other parameters fixed. By holding 
$\omega_{\rm m}$, $\omega_{\rm b}$, and $\theta_{\rm s}$ unchanged, 
one essentially fixes the comoving angular diameter distance to the 
last scattering surface $D_{\rm A}^*$. The effect is that the 
fractional changes of the absolute comoving angular diameter distance 
$D_{\rm A}$ (lines in the left panel) vanish at $z \sim 1100$ and that 
$\Delta \ln D_{\rm A} = \Delta \ln h$ at $z\sim 0$. 
One can measure the absolute distance with standard rulers such as 
BAOs. Since $D_{\rm A}$ is more sensitive to the mean curvature than 
to the dark energy EOS parameters, low-$z$ BAO results could 
have strong 
dependence on the prior of $\Omega_{\rm K}$. As Fig.~\ref{fig:dddg}
suggests, high-redshift distances are efficient for measuring the 
mean curvature accurately as long as dark energy is subdominant
at high redshift \cite{knox06a}, which is indeed demonstrated in 
Ref.~\cite{knox06c} for both photometric and spectroscopic BAO surveys.
Therefore, it is helpful to extend BAO surveys to high redshift.

The behavior of $\Delta \ln D_{\rm A}$ explains the ``geometric
degeneracy'' between the mean curvature and Hubble constant for 
CMB results \cite{spergel06}. With a tight
constraint on $D_{\rm A}^*$ from CMB and assuming $w=-1$, one then 
only needs an accurate measurement of the Hubble constant to determine 
the mean curvature. For example, with $w$ fixed to $-1$, 
$\sigma(\Omega_{\rm K}) \sim 
|\sigma(\ln D_{\rm A})\,{\rm d}\,\Omega_{\rm K}/{\rm d}\ln D_{\rm A}
|_{z\sim 0} \sim 0.12 \sigma(\ln h)$. 
The current error of the Hubble 
constant is at 10\% level \cite{freedman01}, so we can infer that 
$\sigma(\Omega_{\rm K})\sim 10^{-2}$ with CMB and $H_0$ data alone,
assuming $w=-1$.

Type Ia SNe measure the luminosity distance, which is 
the same as the angular diameter distance in comoving coordinates.
Since the absolute luminosity of a SN is degenerate with the Hubble
constant, SN distance is relative, i.e., one determines 
$D_{\rm A}h$ instead of $D_{\rm A}$. The fractional changes of 
$D_{\rm A} h$ are given as symbols in the left panel of 
Fig.~\ref{fig:dddg}. One sees that the parameter sensitivity is 
shifted to higher redshift, which, counter-intuitively, leads to 
measuring the Hubble constant $H_0$ with high redshift data 
\cite{hu05}. This is possible since the error in $D_{\rm A}^* h$
is mostly from $h$. 
In practice, Ref.~\cite{aldering06} finds that high redshift 
($3>z>1.7$) SNe will not dramatically improve dark energy 
constraints, because with the extra data come new observational 
systematics, e.g., one may not even be able to obtain 
redshifts spectroscopically for these SNe. 

The right panel of Fig.~\ref{fig:dddg} shows that although
the linear growth function (lines) is generally less sensitive 
to the dark energy EOS parameters and mean curvature than distance, 
nonlinear evolution (symbols) produces a marked amplification of 
the sensitivity \cite[also noted in Ref.][]{simpson05}. The growth
function is more sensitive to the mean curvature than to the 
dark energy EOS parameters, and the sensitivity peaks at $z \sim 0$.
As such, low redshift measurements of the growth function
can be useful for constraining the mean curvature. 
One may also notice that the growth derivatives with respect to 
$\Omega_{\rm K}$, $w_0$, and $w_a$ have the same sign, whereas the 
distance derivatives do not. This breaks the degeneracy 
between the dark energy EOS parameters and mean curvature and is 
an advantage of being able to measure both distance and
growth well. 

\begin{figure}
\centering
\includegraphics[height=2.835in]{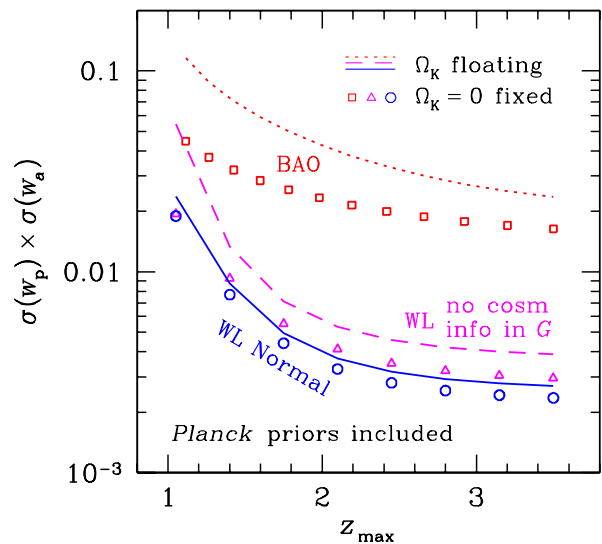}
\caption[f2]{Error product $\sigma(w_{\rm p})\times\sigma(w_a)$ as 
a function of the maximum (photometric) redshift of the data for
BAO (dotted line and open square) and WL (solid line and open 
circles) with (symbols) and without (lines) the flatness assumption.
The results are obtained by discarding BAO or WL bins above 
$z_{\rm max}$. 
In addition, we include the results of a test (dashed
line and open triangles) in which we fix the growth rate to 
the values it has in our fiducial model.
\label{fig:epz}}
\end{figure}

Figure~\ref{fig:epz} gives a demonstration of the importance of 
high-redshift distance measurements for BAO and low-redshift 
data for WL. 
We assume a half-sky photometric BAO and WL survey loosely
based on the proposed Large Synoptic survey Telescope\footnote{
See \url{http://www.lsst.org}.} (LSST) project.
We use 10 WL bins over the photometric redshift range 
$0 \le z_{\rm p} \le 3.5$  and 20 BAO bins over 
$0.15 \le z_{\rm p} \le 3.5$. The widths of the WL bins are equal,
while those of the BAO bins are proportional to $(1+z_{\rm p})$.
The linear galaxy bias is linearly interpolated over 20 bias 
parameters evenly distributed over true redshift $0 \le z \le 4$.
The fiducial model for the galaxy bias is $b(z) = 1 + 0.84 z$, and a
20\% prior is applied to each bias parameter.
The galaxy distribution is proportional to $z^2 \exp(-z/0.5)$ with a 
projected number density of 50 galaxies per square arcmin. The rms 
shear of the galaxies is taken to be 
$\gamma_{\rm rms} = 0.18 + 0.042z$. To isolate the effects under 
investigation, we assume zero uncertainty in the error 
distribution of photometric 
redshifts which we take to be Gaussian with rms $\sigma_z = 0.05(1+z)$ and bias
$\delta z = 0$. The multipoles are limited to $40 \le \ell \le 3000$
for BAO [with an additional requirement that the dimensionless
matter power spectrum $\Delta_\delta^2(k,z) < 0.4$] and 
$40 \le \ell \le 2000$ for WL.
The calculation is performed with the forecasting tool {\sc cswab}
\cite{zhan06d}. 

We use the error product (EP), 
$\sigma(w_{\rm p})\times\sigma(w_a)$, to assess dark energy 
constraints. This product is proportional to the area of the error 
ellipse in the $w_0$--$w_a$ plane, and the error 
$\sigma(w_{\rm p})$ is equal to that on 
$w_0$ with $w_a$ held fixed
\cite{huterer01,hu04b,albrecht06}.

Because of the degeneracy between the galaxy bias and the growth
of the large-scale structure, the BAO technique cannot measure the 
latter accurately even with the redshift distortion 
information in spectroscopic BAO data. This does not contradict the
finding that one can determine the linear galaxy bias to several 
percent with BAO and CMB \cite{zhan06d}. One 
might infer that the growth rate too could be determined to several 
percent as it is degenerate with the galaxy bias. However, the
tight constraints on the linear galaxy bias are obtained 
with assumed cosmological dependence of the distance and 
growth function. If one models the distance and growth rate as
independent free parameters, then the latter (or the product of the 
galaxy bias and growth rate) is poorly determined with BAO
\cite{zhan06c}.

Photometric BAO (dotted line and open squares in Fig.~\ref{fig:epz}), 
being only able to measure the absolute distance (and not the growth),
is susceptible to the prior on the mean curvature, even though it is 
capable of determining $\Omega_{\rm K}$ to $10^{-3}$ level with the 
full range of data \cite{knox06c}. 
This susceptibility is, however, much milder than that of 
relative distances from $z < 1.7$ 
SNe reported in Refs.~\cite{linder05b,knox06c}.
The degradation to the SN constraints due to relaxing the flatness
assumption occurs mostly to $w_a$, because 
the redshift dependence of the derivative 
${\rm d} \ln(D_{\rm A} h)/{\rm d} w_a$ is nearly proportional to
${\rm d} \ln(D_{\rm A} h)/{\rm d} \Omega_{\rm K}$ 
(see Fig.~\ref{fig:dddg}). For BAO both
$w_0$ and $w_a$ are degraded, but the EP is degraded by much less
than in the SN case.

The WL constraints (solid line and open circles in Fig.~\ref{fig:epz}) 
on the dark energy EOS are remarkably robust against
the flatness assumption, despite that the WL error on 
$\Omega_{\rm K}$ is roughly twice that of BAO for LSST 
\cite{knox06c,zhan06d}.
To understand this behavior, we devise a test (dashed line and open
triangles) in which we 
fix growth to the values it has in our fiducial model.
In this case we can, on the one hand, reconstruct distance more 
accurately due to lack of any degeneracy with growth.  On the other 
hand we lose any information about dark energy that comes through the 
dependence of the growth factor on dark energy.  
As expected, the dark energy constraints with only the distance 
information develop some sensitivity to the flatness assumption in 
this case. When the curvature is fixed to 0, the result of this 
test is fairly close to the normal WL ones especially at low 
redshifts where the growth rate is most sensitive to curvature.
This shows that the (low-$z$) growth rate is helpful 
in reducing the error on curvature and, hence, improves the 
constraints on dark energy. However, it also illustrates that the 
power of WL comes mostly from the distance information.

A minor feature of the WL EP in Fig.~\ref{fig:epz} is that 
$z \gtrsim 2$ bins do not contribute much to the dark energy 
constraints, which is also seen for the mean curvature in
Ref.~\cite{knox06c}. This is mainly due to three factors. Firstly, the 
sensitivity of growth to the dark energy EOS and mean curvature 
peaks at low redshift. The peak of the lensing kernel moves away 
from the redshift range that is most sensitive to the parameters when 
the source galaxies are above $z \sim 2$. Secondly, WL is much more 
limited by shot noise than BAO \cite{zhan06d}, so that the poorer 
measurements at higher redshift contribute little to parameter 
constraints. 
If the shape noise were negligible for WL, the 
bins at $z \gtrsim 2$ would continue to improve the EP for bins at
$z \lesssim 2$ by a factor of 2.
In other words, one could see more improvement of WL 
constraints on dark energy at $z_{\rm max} \gtrsim 2$ with a deeper 
survey.
Thirdly, we assume implicitly that dark energy is dominant only at
low redshift with the parametrization $w_0$ and $w_a$ (at the 
fiducial model $w_0 = -1$ and $w_a = 0$). If we allow more degrees 
of freedom in the dark energy EOS, e.g., by modeling it with 
9 parameters evenly spaced between the expansion factor $a = 0.2$ 
and $1$ \cite{albrecht06}, then high-$z$ data will certainly be 
needed to constrain high-$z$ dark energy.

\subsection{Reconstructing Distance and Growth}

We give a pedagogical explanation of how distance and growth
are reconstructed from WL data in this subsection. For simplicity, 
we use a scale-free matter power spectrum with
${\rm d} \ln P_\delta(k)/{\rm d}\ln k = -2.5$ as an example. This is a 
good approximation since the WL technique does not rely on power 
spectrum features as we show in \S \ref{sec:test}.

We begin by reviewing the dependence of the shear power spectra from
two shear maps labeled by $i$ and $j$ 
and the gravitational potential power spectrum $\Delta_\Phi^2$.
For simplicity
we assume all the source galaxies from which shear map $i$ is derived
are at the same redshift and distance, $D_i = D(z_i)$ and similarly
for $j$ (we drop the subscript A for convenience).  
In this case, and using the Limber approximation 
\cite{limber54,kaiser92} the 2-point function for the shear E modes 
in a flat universe is given by \cite{bartelmann01,hu99b}
\bea
\label{eqn:shearint}
P_{ij}^{\gamma\gamma}(\ell) & = &\int dD \frac{D_i-D}{D_i}
\frac{D_j-D}{D_j}\left[k\Delta_\Phi^2(k,z)\right]_{k=\ell/D} 
\nonumber \\ &\times & \Theta(D_i-D)\Theta(D_j-D),
\eea
where $z$ in square brackets corresponds to the redshift
of the comoving angular diameter distance $D(z)$, 
and $\Theta(x)$ is the Heaviside step function.
 
The potential power spectrum is related to the matter power spectrum by $k\Delta_\Phi^2(k) = 
\frac{9}{8\pi^2}\left(\Omega_m H_0^2\right)^2/a^2P_\delta(k)$. Although $P_\delta(k)$ is
not a power law (this is the evolved matter power spectrum at late times, not the primordial 
power spectrum) insight can be gained by considering the power law case.
If $P_\delta(k,z) = Ak^ng_\Phi^2(D)/a^2$, where 
$g_\Phi(D) = G(z) / (1+z)$ is the gravitational potential growth 
factor at the redshift corresponding to the comoving angular distance 
$D(z)$, then
\bea
P_{ij}^{\gamma\gamma}(\ell) &= & A' \ell^n \int \frac{dD}{D} 
\frac{D_i-D}{D_i}
\frac{D_j-D}{D_j}D^{1-n}g_\Phi^2(D) \nonumber \\
&\times & \Theta(D_i-D)\Theta(D_j-D),
\eea
where $A'$ is an easily derived constant.  

We see that in the power law power spectrum case, there is no information about geometry
in the shape of the shear power spectra since the shape only depends on the spectral
index.  Any information about geometry is in the amplitudes of the various shear power 
spectra, which are also affected by the growth function.  Therefore we can count
parameters and degrees of freedom to gain an understanding of the reconstruction of
distance and growth.  If we assume $m$ shear maps and parameterize both the $D(z)$ function
and the $g_\Phi(z)$ function with $m$ parameters each, as was done in Ref.~\cite{knox06b}, then
there are $m(m+1)/2$ knowns (the amplitudes of each of the shear power spectra)
and $2m$ variables to solve for.  For  $m > 2$ the known quantities are more numerous
than the unknown and generally there is a unique 
least-square 
solution.  For a reconstruction
of the parameters of a physical model from several source redshift bins, typically
the number of knowns exceeds the number of parameters.  

Note that if we were to consider $g_\Phi(D)$ to be a completely free function, then
knowledge of absolute distances would be impossible to acquire since a rescaling of
distance would be exactly degenerate with a rescaling of $g_\Phi(D)$. Also, for the special
case of $n=1$ (which is only a good approximation at $\ell \lesssim 30$) there is no
sensitivity to absolute distances even with $g_\Phi(D)$ perfectly known.  Fortunately,
$n=-2.5$ is a better approximation to the shear power spectrum over most of the
$\ell$-range of interest, providing a strong sensitivity to the absolute distance scale
when $g_\Phi(D)$ is parameterized with just a few numbers.  

We can further simplify things by replacing the integral over 
distance in Eq.~(\ref{eqn:shearint}) for the auto shear power 
spectra with the integrand evaluated at an effective lens redshift 
$z_{\rm l}$ with width $\Delta D_{\rm l}$, so that
\be \label{eq:sps}
P_{ii}^{\gamma\gamma}({\ell}) \propto \Delta D_{\rm l} 
D_{{\rm l}i}^2 D_i^{-2} P_\delta(k,z_{\rm l}) \propto
D_{{\rm l}i}^2 D_i^{-2}D_{\rm l}^{2.5} G_{\rm l}^2,
\ee
where $D_{{\rm l}i}$ is the comoving angular diameter distance 
between the lens and the source bin $i$, 
$G_{\rm l} = G(z_{\rm l})$, and 
$P_\delta(k,z_{\rm l}) \propto G_{\rm l}^2 k^{-2.5}$. 

If one models $D_i$, $D_{{\rm l}i}$, $D_{\rm l}$, and $G_{\rm l}$
as independent parameters, then these parameters will have roughly 
the same fractional error because of the similar magnitude of 
exponents. The errors of these parameters will be highly correlated. 
Since a lens distance for one WL bin can become a source distance for
another and vice versa, the same distance is measured by multiple (auto
and cross) shear power spectra. The degenerate directions of the 
distances for different shear power spectra can be quite 
complementary to each other. Therefore, the tomographic WL 
reconstruction of distance is more accurate than that of growth.
The degeneracy may be partially lifted by the 
cosmological model as well.
For example, the distance $D_{{\rm l}i} = D_i - D_{\rm l}$
in a flat universe.

The reconstructed distances (and growth rates) from WL are still 
highly correlated,
but the degeneracies are along directions that cannot be easily 
mimicked by cosmological parameters. This is why even though the 
distances reconstructed with LSST BAO \cite{zhan06c} have smaller 
marginalized errors than those with LSST WL \cite{knox06b}, the
dark energy constraints from the former is nevertheless weaker 
(note that the growth information itself contributes only in a 
minor way, see Fig.~\ref{fig:epz}).

For comparison, the angular galaxy power spectra with the same 
approximation are
\be \label{eq:gps}
P_{ii}^{\rm gg}(\ell) \propto D_i^{0.5} b_i^2 G_i^2,
\ee
where $b_i$ is the linear galaxy bias. Since the galaxy power 
spectra, unlike WL shear power spectra, are local to the 
underlying galaxy distribution, i.e., 
no mixing of distances at other redshifts in Eq.~(\ref{eq:gps}), 
one cannot measure the same distance multiple 
times with different auto and cross galaxy power spectra to break 
the degeneracies and reduce the errors. The exponents of 
$D_i$, $b_i$, and $G_i$ in Eq.~(\ref{eq:gps}) suggest
that, with distances measured from the BAO feature,
one can then infer from $P_{ii}^{\rm gg}$ the product $b_i G_i$ 
with one fourth the fractional distance error.  In the case
that growth is constrained tightly from (a combination with) other 
data (WL for example) and a physical model,
then the bias parameter can be determined quite precisely as was
seen in Refs.~\cite{hu04b,zhan06d}.

\begin{figure*}
\centering
\includegraphics[height=2.835in]{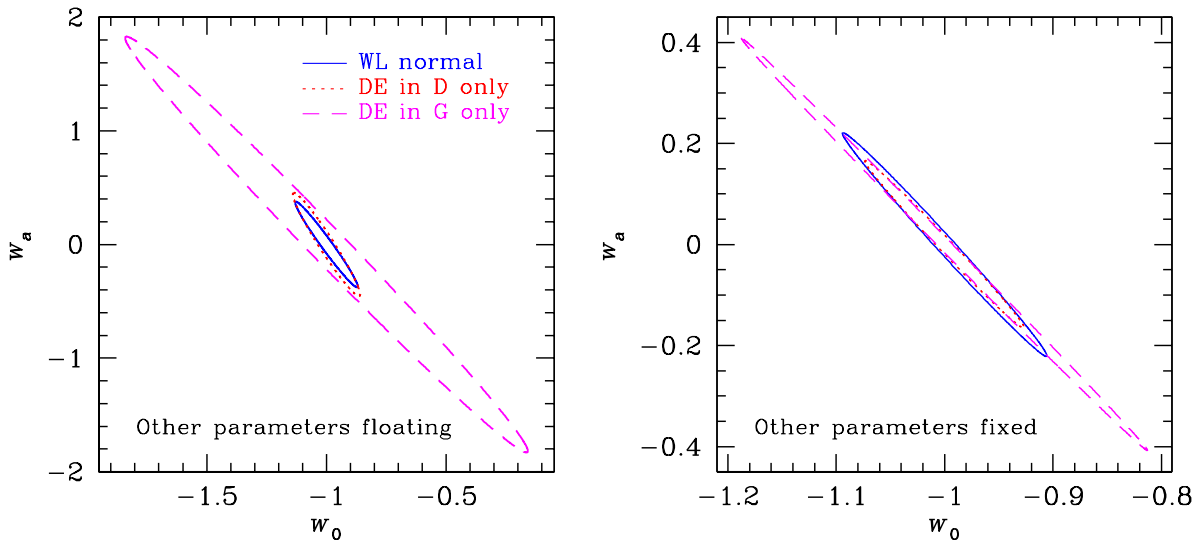}
\caption[f3]{\emph{Left panel}: $1\sigma$ error contours in the 
$w_0$--$w_a$ plane for normal WL (solid lines), WL with dark 
energy information in the growth rate discarded (dotted line), and
WL with dark energy information in the distance discarded
(dashed line). The results are marginalized over all the other 
cosmological parameters. \emph{Right panel}: Same as the left but 
with all the other parameters fixed. The distance-only contour is 
slightly smaller than the normal WL contour in this panel because
of the partial cancellation between the distance and growth effects
\cite{simpson05}.
\label{fig:dgde}}
\end{figure*}

\subsection{Dark Energy Constraints from Distance and Growth}

We examine the relative strength of dark energy constraints 
from the distance and growth rate with the WL technique. 
In the same spirit as the test in Fig.~\ref{fig:epz}, we 
evaluate the dark energy constraints from the 
distance by not allowing the growth to depend on  
dark energy parameter variations
(``DE in D only'') and vice versa (``DE in G only'').  
We take independent Gaussian priors for the tests:
$\sigma_{\rm P}(\ln\omega_{\rm m})=\sigma_{\rm P}(\ln\omega_{\rm b})=
\sigma_{\rm P}(n_{\rm s})=\sigma_{\rm P}(\alpha_{\rm s}) = 
\sigma_{\rm P}(\ln\Delta_R^2)=0.05$,
$\sigma_{\rm P}(\theta_{\rm s})=0.003^\circ$, 
$\sigma_{\rm P}(\tau)=0.01$, and
$\sigma_{\rm P}(Y_{\rm p})=0.02$. These priors are fairly 
conservative compared to what can be achieved with {\it Planck},
and they do not introduce extra correlations between the parameters.

The results are shown in Fig.~\ref{fig:dgde}. When
marginalized over all the other cosmological parameters (left panel),
the distance (dotted line) is far more powerful than the growth rate
(dashed line), and the WL constraints on dark energy (solid line) are
almost entirely from the distance. This is consistent with the 
conclusions in Refs.~\cite{abazajian03,knox06b}. The 
difference between the constraints from the distance and growth rate
depends on the priors on the other parameters. If one fixes
all the other parameters (right panel), then the distance and growth
rate offer comparable dark energy constraints, in 
agreement with Refs.~\cite{simpson05,zhang05}. 

Note that, as can be seen in Fig.~\ref{fig:dgde}, the constraints on
dark energy are tighter in the ``DE in D only'' case than in the ``WL
normal'' case.  This is because there are two effects of fixing $G$
with respect to dark energy variations.
One effect is that one now has better distance information because
there is no degeneracy with growth.  The other, apparently subdominant
effect, is the loss of any information that comes through the
dependence of growth on dark energy.  
The improvement in the
artificial case of ignored growth dark energy dependence 
is also consistent with the notion that there is a partial
cancellation of the effects on the shear power spectra of varying 
distance and growth through the dark energy EOS \cite{simpson05}.

\subsection{Why Does WL Perform Better than BAO?}

We now get at the question of how WL surveys constrain dark energy by
comparing them with BAO surveys.  There are two major differences
between the two probes.  First, there is an unknown bias that affects
the galaxy power spectra.  The bias means that growth
information is not accessible, since growth and bias are degenerate.
Also due to the bias, analysis of galaxy power spectra is typically
restricted to mostly linear scales to avoid the complications of a
scale-dependent bias (besides nonlinear evolution of the matter 
power spectrum itself).  Second, the kernels are different.  The WL
kernel is broader, suppressing sensitivity to matter power spectrum
features, and it also depends on distances, making the amplitude of
the shear power spectra sensitive to the distance-redshift relation.

We can gain insight by comparing a WL survey to something
even better than a galaxy survey:  a map of the dark matter
density as a function of redshift and angular coordinates,
with no peculiar velocity contributions to the redshifts.  
Would such a survey be better than a WL survey at determining
dark energy parameters?  One might think so since the shear
maps are merely projections over this density field, projections
that will suppress the prominence of power spectrum features.  However, our
tests in the next section show that the WL survey is significantly 
more constraining.  The reason is that the projection is dependent 
on the distance-redshift relation and, in fact, is the main source 
of sensitivity  of the shear power spectra to the distance-redshift relation.

Thus our conclusion is that while the lack of any unknown bias
factor may be important, it is not the crucial difference between WL
and galaxy surveys.  The WL kernel is the most important difference.

The bias factor is indeed an important difference as well.  If WL were
affected by an unknown bias factor that was a free function of
redshift, then the constraints on dark energy would be greatly
degraded.  In apparent contradiction we have also seen that the constraints
on dark energy from the growth information are highly subdominant
to those from distance information.  These two statements
are both true.  The reason an unknown bias factor would degrade the
dark energy constraints is not due to the loss of growth information,
but because without priors on the growth of the gravitational potential
as a function of redshift, the distance reconstruction degrades.  

In fact, if we let growth be a free function, this would be the same
as letting the bias be a free function.  The quality of the distance
reconstruction from WL depends on the growth-redshift relation having
a sufficiently small number of parameters --- whether those are the
parameters of the cosmological model or the parameters of a phenomenological
model such as the growth factor evaluated at a discrete set of points
in redshift space, as was done in Refs.~\cite{song05} and \cite{knox06b} 
and as was discussed above.    

Finally, we should be clear that although WL gets tighter constraints
on the distance-redshift relation than BAO, this is not necessarily
apparent from looking at the errors on the distances to particular
redshifts.  The BAO ones can be smaller.  However, the errors in the
distances to particular redshifts in the WL case are highly correlated
with the errors in the distances to other redshifts.  Because of these
correlations, there are linear combinations of distances with much
smaller errors than any individual distance.  
These strong
correlations are present because any single auto or cross shear power
spectrum depends on the distance to many redshifts.  

\begin{table}
\caption{Convention of test names.\label{tab:name}}
\begin{ruledtabular}
\begin{tabular}{llll}
Prefix & Feature & Suffix & Feature
\\[0.4ex] \hline \\[-1.5ex]
SF & scale-free\footnotemark[1]  & LE & linear evolution \\
NW & no wiggle\footnotemark[2]   & NE & nonlinear evolution  \\
WW & with wiggle\footnotemark[3] & SR & artificial standard ruler \\
   &             & AB & artificial bias for WL \\
   &             & FB & fixed bias for BAO \\
\end{tabular}
\end{ruledtabular}
\footnotetext[1]{With the spectral index 
${\rm d}\ln P(k)/{\rm d} \ln k = -2.5$ and amplitude 
proportional to that of the CDM power spectrum. }
\footnotetext[2]{Given by the fitting formula in 
Ref.~\cite{eisenstein99a}.}
\footnotetext[3]{Calculated with {\sc cmbfast} \cite{zaldarriaga00}.}
\end{table}

\section{Power Spectrum Features and Cosmological
Constraints} \label{sec:test}

We devise a number of tests to explore the impact of power spectrum
features, such as the broadband turnover, BAOs, and nonlinear 
evolution, on the constraints of dark energy and related cosmological
parameters. The tests are divided into two broad categories: with and
without the shot noise. The former has the advantage that the 
overall amplitude of the signal, i.e., the normalization of the galaxy 
or shear power spectra, does not affect the Fisher matrix for error 
estimation. In other words, only the power spectrum shapes and 
their relative amplitudes affect the constraints when the shot
noise is neglected. Clearly, the noise-free case is unrealistic. 
The inclusion of the shot noise degrades the results (more severely 
for WL), and in this case a boost to the power spectrum and its 
sensitivity to parameters due to nonlinear evolution can be helpful.

We use a scale-free matter power spectrum, a CDM power spectrum 
with no baryon wiggles, and a CDM power spectrum with wiggles
to carry out the tests. For each type, we specify up to 5 
additional characteristics: linear evolution, nonlinear 
evolution, artificial standard ruler on nonlinear scales, 
artificial bias 
for WL, and fixed galaxy bias for BAO. Table~\ref{tab:name}
summarizes the convention of test names; for example, NWLE refers
to the no-wiggle CDM power spectrum that evolves according to the
linear theory on all scales.

Note that linear evolution applies to all cases unless nonlinear 
evolution is specified. For SFNE and WWNE, the relative boost to
the matter power spectrum due to nonlinear evolution is set to be 
the same as that of NWNE, 
in order to avoid the difficulty of applying the \citet{peacock96}
fitting formula for the nonlinear power spectrum to the CDM power 
spectrum with wiggles \cite{zhan06d}. In addition, using the same
relative boost of the power spectrum ensures the same parameter 
sensitivity arising from the nonlinear correction.

The artificial standard ruler is given by 
\be \label{eq:asr}
\frac{P_{\rm SR}(k)}{P_\delta(k)} = 1+9\left(1- e^{-u^2}\right) +
9 \sqrt{q}\left(1-e^{-v^2}\right),
\ee
where $q = k / (h\,\mbox{Mpc}^{-1})$, $u = q/1.5$, and 
$v = q / 3$. This feature mimics the 
\citet{peacock96} fitting formula of the nonlinear matter power 
spectrum for our fiducial cosmological model at $z = 0$. Its 
independence of redshift makes the artificial feature
standard and, meanwhile, stronger than nonlinear evolution 
at $z > 0$. Since we impose the condition 
$\Delta_\delta^2(k,z) < 0.4$ and $40 \le \ell \le 3000$ for galaxy 
power spectra, the artificial standard ruler and nonlinear 
evolution affect our analysis only through their influence
on larger scales. The artificial bias for WL is 
implemented in the same
way as the galaxy bias but with a fiducial model of $b = 1$. 
A 20\% prior on the galaxy and artificial bias parameters is 
always applied.

\begin{table}
\caption{Parameter errors from noise-free galaxy power spectra. 
\label{tab:baonf}}
\begin{ruledtabular}
\begin{tabular}{lddddd}
 & & & \multicolumn{1}{c}{$\ln \omega_{\rm m}$} & 
\multicolumn{1}{c}{$\Omega_{\rm K}$} & 
\multicolumn{1}{c}{$\ln \Delta_R^2$} 
\\[0.4ex] \cline{4-6} \\[-1.8ex]
\text{Case} & \multicolumn{1}{c}{$w_{\rm p}$} & 
\multicolumn{1}{c}{$w_a$} & \multicolumn{3}{c}{($10^{-2}$)} 
\\[0.4ex] \hline \\[-1.5ex]
SFLE & 2.3   & \multicolumn{1}{c}{$14$}   & 5.0 & 
       \multicolumn{1}{c}{$13$}   & 5.0 \\
SFNE & 0.18  & 1.6  & 2.2 & 1.3  & 4.9 \\
SFSR & 0.037 & 0.81 & 4.6 & 0.98 & 4.9 \\
SFFB & 0.032 & 0.25 & 2.6 & 0.41 & 4.3 \\
SFFB\footnotemark[1] & 0.019 & 0.13 & 2.6 & 0.37 & 4.3 \\[0.4ex]
\hline \\[-1.5ex]
NWLE & 0.11  & 1.5 & 3.5 & 0.95 & 4.8 \\
NWNE & 0.11  & 1.4 & 3.6 & 0.61 & 4.8 \\
NWSR & 0.037 & 0.85    & 3.8 & 0.87 & 4.8 \\
NWFB & 0.018 & 0.12    & 3.3 & 0.38 & 2.1 \\
NWFB\footnotemark[1] & 0.012 & 0.10    & 2.8 & 0.38 & 1.3 \\[0.4ex]
\hline \\[-1.5ex]
WWLE & 0.034       & 0.58    & 2.7 & 0.22 & 4.8 \\
WWNE & 0.035       & 0.51    & 2.2 & 0.21 & 4.8 \\
WWFB & 0.012 & 0.08 & 2.4 & 0.20 & 1.9 \\
WWFB\footnotemark[1] & 0.010 & 0.07 & 2.5 & 0.25 & 1.9 \\
\end{tabular}
\end{ruledtabular}
\footnotetext[1]{These FB results are calculated in the same way 
as the WL ones in Table~\ref{tab:wlnf}, i.e., with 10 uniform bins 
and $40 \le \ell \le 2000$ without a limit on $\Delta_\delta^2(k,z)$.}
\end{table}

\subsection{Noise-Free Case}

Since BAO and WL are not sensitive enough to all the parameters
(e.g., the electron optical depth of the reionized intergalactic
medium), at least some parameter priors are needed to regularize the
constraints \cite{song04}. 
The correlation of parameters in CMB priors can obscure the effect 
of power spectrum features we try to isolate. Hence, in the 
noise-free case we take the same independent Gaussian priors as
those in Fig.~\ref{fig:dgde}.

Table~\ref{tab:baonf} presents the marginalized errors on $w_{\rm p}$,
$w_a$, $\omega_{\rm m}$, $\Omega_{\rm K}$, and $\Delta_R^2$
from noise-free angular galaxy power spectra with various 
underlying matter power spectra. As we expect, SFLE does not
provide meaningful constraints on the parameters, because it has
no feature for measuring the absolute distance and because the 
galaxy bias prevents one from extracting the distance and 
growth information from the amplitude of the galaxy power spectra. 
The redshift-dependent, but predictable, ruler provided by 
nonlinear evolution leads to an improvement of SFNE over SFLE.  
However, the length of this ruler cannot be predicted very 
accurately since it depends on 
both the shape and amplitude of the matter power spectrum.
Therefore when other rulers are present, the
addition of nonlinear evolution makes very little difference,
as in the case of NWNE vs. NWLE and WWNE vs. WWLE.
We should point out that the tests in Table~\ref{tab:baonf} 
do not take full advantage of the nonlinear evolution as we 
truncate the matter power spectrum at largely linear scales.
Extending the analysis to smaller scales does increase the 
difference between WWLE and WWNE results (see \S~\ref{sec:withsn}),
but doing so requires one to model the scale-dependent galaxy
bias on those scales very accurately.

The artificial standard ruler of Eq.~(\ref{eq:asr}) 
is similar to the nonlinear feature at $z = 0$ but
independent of redshift. It offers a great improvement on distance
measurements over SFNE, which is evident from the errors on 
$w_{\rm p}$, $w_a$, and $\Omega_{\rm K}$, but, having nothing to
do with $\omega_{\rm m}$, it does not reduce 
$\sigma(\ln\omega_{\rm m})$. When the galaxy bias is fixed in
SFFB, the distance and growth can be measured through the 
amplitudes of the galaxy power spectra, 
which have distance dependencies
and lead to the smallest errors
on nearly all parameters within the SF group. 
One still cannot do well on $\Delta_R^2$ with SFFB, because the 
normalization of growth and that of distance  are 
degenerate for scale-free galaxy power spectra.
The sensitivity of SFFB to $\omega_{\rm m}$ comes from our choice 
of normalizing the matter power spectrum to the CMB potential 
fluctuations \cite{zhan06d}.

\begin{table}
\caption{Parameter errors from noise-free shear
power spectra. \label{tab:wlnf}}
\begin{ruledtabular}
\begin{tabular}{lddddd}
 & & & \multicolumn{1}{c}{$\ln \omega_{\rm m}$} & 
\multicolumn{1}{c}{$\Omega_{\rm K}$} & 
\multicolumn{1}{c}{$\ln \Delta_R^2$} 
\\[0.4ex] \cline{4-6} \\[-1.8ex]
\text{Case} & \multicolumn{1}{c}{$w_{\rm p}$} & 
\multicolumn{1}{c}{$w_a$} & \multicolumn{3}{c}{($10^{-2}$)} 
\\[0.4ex] \hline \\[-1.5ex]
SFLE & 0.0032    & 0.023    & 1.9 & 0.19 & 3.5 \\
SFNE & 0.0034    & 0.027    & 1.5 & 0.14 & 2.7 \\
SFSR & 0.0031    & 0.022    & 1.9 & 0.18 & 3.4 \\
SFAB & 0.014 & 0.16 & 4.2 & 0.56 & 4.7 \\[0.4ex]
\hline \\[-1.5ex]
NWLE & 0.0036    & 0.021    & 2.1 & 0.17 & 1.1 \\
NWNE & 0.0035    & 0.026    & 2.1 & 0.14 & 0.37 \\
NWSR & 0.0030    & 0.022    & 2.0 & 0.16 & 1.1 \\
NWAB & 0.013 & 0.13 & 2.5 & 0.50 & 4.5 \\[0.4ex]
\hline \\[-1.5ex]
WWLE & 0.0033 & 0.020 & 1.9 & 0.16 & 1.3 \\
WWNE & 0.0033 & 0.025 & 1.2 & 0.13 & 0.78 \\
WWAB & 0.0089 & 0.072 & 2.6 & 0.30 & 4.5 \\
\end{tabular}
\end{ruledtabular}
\end{table}

With the standard ruler of the broadband turnover in the matter 
power spectrum, NWLE performs much better than SFLE, and nonlinear
evolution becomes a subdominant factor. As in SFSR, the artificial 
standard ruler in NWSR offers yet better measurements of the absolute
distance and stronger constraints on parameters. Because of the 
broadband turnover, a shift in the absolute distance will not only 
alter the amplitude of the angular galaxy power spectra but also
move the broadband feature in multipole space. This breaks the 
degeneracy between the normalizations of growth and distance seen in 
SFFB, so that NWFB improves $\Delta_R^2$ (in addition to other 
parameters) significantly.

By comparing the results of the NW tests and those of the WW tests,
one sees that the BAOs are by far the most crucial feature for 
measuring the absolute distance and constraining cosmology with
galaxy power spectra. Their importance relative to that of the
broad band feature was already convincingly demonstrated by
Ref.~\cite{seo03}.  Also, WWFB demonstrates that it 
would be highly advantageous to be able to predict the bias
to high precision since the results improve dramatically
for fixed galaxy bias. 

Table~\ref{tab:wlnf} gives the results from the noise-free WL 
shear power spectra. 
The most striking characteristic is that these results 
are fairly independent of all our modeling changes, except for
the introduction of artificial bias.  There is no 
substantial difference between the dark energy constraints
from completely featureless SFLE and those from WWNE.
The larger errors on $\Delta_R^2$ with the scale-free matter power 
spectrum is due to the degeneracy between the normalization of
distance and that of growth, as is the case in Table~\ref{tab:baonf}.
This shows that the WL constraints on dark energy are 
primarily derived from the correlated distances, which are
inferred from the amplitudes
of the shear power spectra. One would not obtain the same
distance constraints without the ability to measure the growth 
function. This is illustrated by the severe degradation to the 
results of SFAB, NWAB, and WWAB for which the lensing potential is 
artificially sourced by unknown bias parameters 
\cite[see also][]{knox06c}. 
One may notice among the AB tests that WWAB is considerably better
than the other two as far as $w_0$ and $w_a$ are concerned. This 
reveals that the standard features in the matter power spectrum do 
contribute somewhat to WL constraints on dark energy, but they are 
much less important than the lensing kernel and the lack of bias in 
the WL technique.

It is also of interest to compare the results of SFFB in 
Table~\ref{tab:baonf} with those of SFLE in Table~\ref{tab:wlnf}.
In the noise-free case, the errors of the shear power spectra and 
those of the galaxy power spectra are given identically by the cosmic
variance. 
For the same scale-free matter power spectrum and fixed 
galaxy bias, Eqs.~(\ref{eq:sps}) and (\ref{eq:gps}) predict a
stronger sensitivity of shear power spectra to distances
than is the case for galaxy power spectra.  Thus we expect to have
tighter constraints on distances from shear power spectra than from
galaxy power spectra, although  the former will probably have a more complicated
error correlation structure.   The expectation of tighter constraints
is indeed supported by the much smaller errors on
$w_{\rm p}$ and $w_a$ from SFLE in Table~\ref{tab:wlnf} compared to 
those from SFFB in Table~\ref{tab:baonf}. 

\subsection{Constraints with the shot noise}\label{sec:withsn}

The noise-free tests in the previous sub-section are 
useful for understanding the origins of cosmological constraints 
for the BAO and WL techniques, though the constraints 
can behave quite differently with noise.
To evaluate the effect of nonlinear evolution more realistically,
we include the shot noise and apply CMB priors from {\it Planck}
to the tests in Table~\ref{tab:wbnoi}.

\begin{table}
\caption{\label{tab:wbnoi}Cosmological constraints from galaxy 
and shear power spectra with the shot noise and \textit{Planck} 
priors.} 
\begin{ruledtabular}
\begin{tabular}{lcddddd}
 & & & & \multicolumn{1}{c}{$\ln \omega_{\rm m}$} & 
\multicolumn{1}{c}{$\Omega_{\rm K}$} & 
\multicolumn{1}{c}{$\ln \Delta_R^2$} 
\\[0.4ex] \cline{5-7} \\[-1.8ex]
\text{Case} & \text{Probe} & \multicolumn{1}{c}{$w_{\rm p}$} & 
\multicolumn{1}{c}{$w_a$} & \multicolumn{3}{c}{($10^{-2}$)} 
\\[0.4ex] \hline \\[-1.5ex]
WWLE & BAO & 0.037 & 0.66 & 0.50 & 0.11 & 1.8 \\
WWNE & BAO & 0.040 & 0.59 & 0.51 & 0.10 & 1.8 \\[0.4ex]
\hline \\[-1.5ex]
WWLE\footnotemark[1] & BAO & 0.027 & 0.49 & 0.50 & 0.099 & 1.8 \\
WWNE\footnotemark[1] & BAO & 0.028 & 0.38 & 0.48 & 0.089 & 1.8 \\[0.4ex]
\hline \\[-1.5ex]
WWLE & WL  & 0.019 & 0.23 & 0.84 & 0.22 & 1.8 \\
WWNE & WL  & 0.014 & 0.19 & 0.67 & 0.19 & 1.7 \\
\end{tabular}
\end{ruledtabular}
\footnotetext[1]{These BAO results are obtained with wider range
of scales by relaxing $\Delta_\delta^2(k,z)$ to less than unity
instead of $\Delta_\delta^2(k,z) < 0.4$.}
\end{table}

Compared to the results in Table~\ref{tab:baonf}, the shot noise 
increases the BAO EP, $w_{\rm p}\times w_a$, only mildly.
The improvement to other parameters is due to the {\it Planck}
priors. The inclusion of nonlinear evolution decreases BAO EP by 
merely 2\%, because we only use the galaxy
power spectra on largely linear scales. If we extend the scales of
the analysis by relaxing $\Delta_\delta^2(k,z)$ to less than unity
(instead of $\Delta_\delta^2(k,z) < 0.4$), the improvement on the EP 
due to the nonlinear feature elevates to 20\%. 
However, it may be quite optimistic as we have not 
accounted for our uncertain knowledge of the scale-dependent 
galaxy bias in the quasi-linear regime.

Unlike BAO, the WL technique is more limited by the shot noise, 
so that the results in Table~\ref{tab:wbnoi} are a lot worse than
those in Table~\ref{tab:wlnf}. In other words, the WL constraints 
should be sensitive to nonlinear evolution, which boosts the 
shear power spectra on scales where the shot noise is dominant. 
Hence, the WWNE WL constraints are considerably tighter than the
WWLE WL constraints in Table~\ref{tab:wbnoi}, 
even though WL results can be degraded by 
nonlinear evolution in the noise-free case.

\section{Conclusions} \label{sec:con}

The dependence of WL shear power spectra on
$D(z)$ and $G(z)$ is much more complicated than, say,
that of supernova luminosity distance.  Here we
have employed a number of numerical tests and analytic
arguments to gain an understanding of {\em how} WL observations
recover information about $D(z)$ and $G(z)$ and thus
constrain cosmology and dark energy in particular.  We
have also similarly explored the cosmological constraints
from the correlations of number densities of galaxies
in photometric redshift bins. 

We have found that, in contrast to the case with galaxy
correlations, standard rulers in the matter power spectrum
play no significant role in $D(z)$ reconstruction from WL surveys.
Instead, the dependence of the lensing kernel on distance
ratios allows for a determination of the distance ratios.
Furthermore, the shear power spectra are directly connected
to the matter power spectrum without any bias, so that one can
determine the degenerate normalizations of distance and growth 
from the amplitude of the shear power spectra. This
degeneracy can be broken by assuming a physical model for either
quantity, which occurs, for example, when one projects the 
constraints on distances and growth rates into cosmological 
parameter space. With WL, the growth information itself is 
less powerful than the distance information in constraining dark 
energy. However, as mentioned in \S~\ref{sec:intr}, it is crucial
to have the ability to measure the amplitude of the matter power 
spectrum. Otherwise, the WL technique would achieve much lower 
precision on the distance-redshift relation and therefore much
lower precision on cosmological parameters as well.

With galaxy power spectra if there were no confusion
from the galaxy bias then knowing the amplitude of the matter power 
spectrum would also help in determining the distance-redshift relation 
from the amplitude of the galaxy power spectra.  We find that
the ruler from nonlinear evolution, 
due to its redshift and model dependencies, is more difficult to 
standardize than other features in the matter power spectrum such 
as the BAOs. 
It will not help with constraints from galaxy power spectra unless
the scale-dependent galaxy bias can be modeled very well.

We expect this understanding of how galaxy surveys constrain cosmology
(via WL in particular) to be useful for analytic consideration of
various effects such as systematic errors, enlarged cosmological
parameter spaces, and the inclusion of complementary information.

\acknowledgments 
We thank J.~A.~Tyson for helpful comments on the manuscript and
A.~J. Hamilton for a useful conversation.
This work was supported 
by NASA grant NAG5-11098 and NSF grants No. 0307961 and 0441072.


\end{document}